\documentclass[a4paper,11pt]{article}
\usepackage{pos}

\title{
Nucleon isovector form factors from domain-wall lattice QCD at the physical mass}

\author*[a]{Shigemi Ohta}

\affiliation[a]{Institute of Particle and Nuclear Studies, High-Energy Accelerator Research Organization (KEK), Tsukuba, Ibaraki 305-0801, Japan}

\emailAdd{shigemi.ohta@kek.jp}

\abstract{
\vspace{-130mm}\parbox{\textwidth}{\flushright\large\rm \hfill KEK-TH-2564}\vspace{123mm}
The current status of lattice-QCD numerical calculations by joint LHP, RBC, and UKQCD collaborations of nucleon isovector vector- and axialvector-current form factors using a 2+1-flavor dynamical domain-wall fermions lattice QCD ensemble generated jointly by RBC and UKQCD collaborations are presented.
The lattice spacing is set at about 0.1141(3) fm, and the lattice spatial extent is 48 spacings or about 5.4750(14) fm. 
The strange and degenerate up and down quark mass values are set at their essentially physical values to provide the physical \(\Omega\) mass and a degenerate pion mass of 0.1392(2) GeV.
Our nucleon mass estimate is about 0.947(6) GeV.
Possible excited-state contaminations in the calculated vector- and axialvector-current form factors are hidden below larger statistical noises.
The numerical details of the form-factor shape parameters, such as the mean squared radii, the anomalous magnetic moment, or the pseudoscalar coupling extracted from the form factors, are described, along with comparisons of different approaches used to extract them.
}

\FullConference{The 40th International Symposium on Lattice Field Theory (Lattice 2023)\\
July 31st - August 4th, 2023\\
Fermi National Accelerator Laboratory\\}

\begin{document}
\maketitle

\section{Introduction}

RIKEN-BNL-Columbia (RBC) and UKQCD collaborations have been jointly studying nucleon structure using dynamical domain-wall fermions (DWF) numerical lattice-QCD ensembles \cite{Lin:2008uz,Yamazaki:2008py,Yamazaki:2009zq,Aoki:2010xg,Abramczyk:2019fnf}.
To extract the nucleon observables, we use the standard ratios,
\(
C^{(3)\Gamma,O}(t_{\rm src},t,t_{\rm snk}) / C^{(2)}(t_{\rm src},t_{\rm snk})
\), of two-point,
\(
C^{(2)}
= \sum_{\alpha\beta} 
\left(
(1+\gamma_t)/2
\right)_{\alpha\beta}
\langle
N_\beta(t_{\rm snk})\bar{N}_\alpha(t_{\rm src})
\rangle,
\)
and three-point,
\(
C^{(3)\Gamma, O}
= \sum_{\alpha\beta}
\Gamma_{\alpha\beta}\\
\langle
N_\beta(t_{\rm sink})O(t)\bar{N}_\alpha(t_{\rm src})
\rangle,
\)
correlators with a nucleon operator, \(N=\epsilon_{abc}(u_a^T C \gamma_5 d_b) u_c\), and an appropriate observable operator \(O\). 
Plateaux of these ratios in time between the source and sink are obtained with appropriate spin (\(\Gamma=(1+\gamma_t)/2\) or \((1+\gamma_t)i\gamma_5\gamma_k/2\)) or momentum-transfer projections, which in turn give lattice bare value estimates for the expected values, \(\langle O\rangle\), for the relevant observables.
Further details can be found in our earlier publications, such as Ref.\ \cite{Yamazaki:2009zq}.

Most recently, the Lattice Hadron Physics (LHP) collaboration joined the effort using a physical-mass ensemble \cite{Syritsyn:2014xwa,Ohta:2015aos,Ohta:2017gzg,Ohta:2018zfp,Ohta:2019tod,Ohta:2021ldu,Ohta:2022csu}.
In Lattice 2022, I reported the nucleon isovector vector- and axialvector-current form factors \cite{Ohta:2022csu} calculated jointly by LHP, RBC, and UKQCD collaborations using the 2+1-flavor dynamical domain-wall fermions lattice-QCD ensemble generated jointly by RBC and UKQCD collaborations.
In this ``48I'' ensemble \cite{RBC:2014ntl}, 
the lattice spacing is set at about 0.1141(3) fm, and the lattice spatial extent is 48 spacings or about 5.4750(14) fm. 
The dynamical strange and degenerate up and down quark mass values are set at their essentially physical values to provide the physical \(\Omega\) mass and a degenerate pion mass of 0.1392(2) GeV.
Our nucleon mass estimate is about 0.947(6) GeV.
Though possible excited-state contamination was detected in the nucleon isovector vector charge, such contamination was not detected in the axialvector charge nor any of the form factors.
Thus, we proceeded to extract shape parameters such as mean-squared radii, the anomalous magnetic moment, and the pseudoscalar coupling, as was summarized in a table that I reproduce here as Table \ref{tab:summary} for the readers' convenience.
\begin{table}[b]
\begin{center}
\begin{tabular}{llllllll}\hline\hline
 & & \(T=8\) & 9 & 10 & 11 & 12 & experiment \\\hline\hline
\(\langle r_1^2\rangle\) & linear&0.134(14) & 0.14(2) & 0.13(3) &  0.16(5) & 0.13(8) & 0.868(3) \(\mbox{fm}^2\)\\
 & dipole & 0.135(6) & 0.143(8) & 0.142(13) & 0.14(2) & 0.13(3) & \\\hline
\(F_2(0)\) & linear & 3.159(4) & 3.250(6) & 3.242(8) & 3.252(13) & 3.61(2) & 3.705874(5)\(\mu_N\) \\
 & dipole & 3.10(5) & 3.15(6) & 3.22(8) & 3.24(11) & 3.5(2) & \\\hline
\(\langle r_A^2\rangle\) & linear & 0.177(2) & 0.174(2) & 0.182(4) & 0.192(5) & 0.066(8) & 0.5(2)\cite{MINERvA:2023avz}\\
 & dipole & 0.177(7) & 0.174(10) & 0.176(14) & 0.18(2) & 0.15(3) & \\\hline
\(F_P(0)\) & linear & 21.01(3) & 22.61(5) & 23.90(7) & 23.04(11) & 26.5(2) & -- \\
 & dipole & 23(2) & 25(2) & 26(2) & 26(2) & 30(2)& \\
\hline\hline
\end{tabular}
\end{center}
\caption{
\label{tab:summary}
The isovector form factor shape parameters obtained by dipole fits agree with those from linear extrapolations using only the smallest two \(Q^2\) values.
The vector-current parameters, however, disagree with well-established experiments \cite{Workman:2022ynf}.
The errors are single-elimination jack-knife statistical. 
}
\end{table}

In that report, I used two methods to extract the shape parameters \cite{Ohta:2022csu}: 1) linear determination using the two smallest momenta transfer values available, and 2) dipole fits to \(\propto (1+Q^2/M_p^2)^{-p}\) with \(p=2\).
Since the two methods broadly agreed with each other, and they do not differ much from fits using other multipolarity \(p = 1, 3, 4, ...7\), I also commented that ``the shape parameter estimates from other fit ansatze, such as bounded-\(z\) expansion, should not differ either, though we are yet to complete such analyses.''
Here, I like to follow up on this comment.

\section{Naive bounded-\(z\) polynomial expansion}

\begin{figure}[t]
\begin{center}
\includegraphics[width=0.45\linewidth,clip]{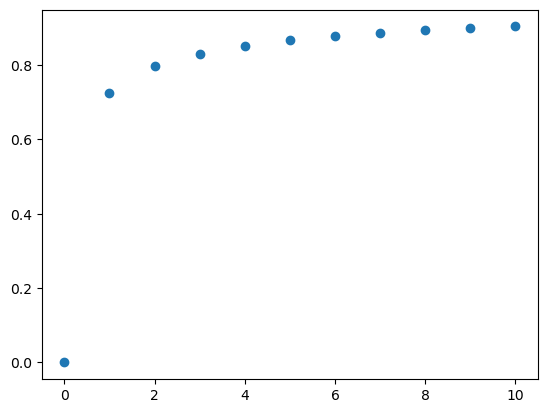}
\includegraphics[width=0.45\linewidth,clip]{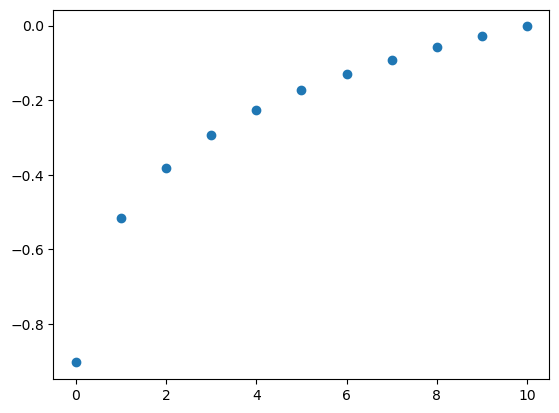}
\end{center}
\caption{
\label{fig:zmaps}
The two \(Q^2 \mapsto z\) mappings we present here, the one with \(t_0 = 0\) (left pane) and the other with \(t_0 = -Q^2_{\rm max} = -10\) (right pane).
}
\end{figure}

Bounded \(z\) parameter
\begin{equation}
z(q^2) = z(t; t_0, t_{\rm cut}) = 
\frac{\sqrt{t_{\rm cut}-t}-\sqrt{t_{\rm cut}-t_0}}{\sqrt{t_{\rm cut}-t}+\sqrt{t_{\rm cut}-t_0}},
\end{equation}
maps \(t = q^2 = -Q^2\) to within the unit disk \(\lvert z \rvert \le 1\).
Form factors can be expanded by polynomials of \(z\),
\begin{equation}
F(Q^2) = \sum_{k=0}^{k_{\rm max}}F_k z(q^2)^k,
\end{equation}
with appropriate \(t_{\rm cut} = 4 m_\pi^2\) for vector and \(t_{\rm cut} = 9 m_\pi^2\) for axialvector currents  \cite{Lee:2015jqa}.
The parameter \(t_0\) allows us to adjust \(Q^2 \mapsto z\) mapping (see Fig. \ref{fig:zmaps}).

\begin{figure}[t]
\begin{center}
\includegraphics[width=0.45\linewidth,clip]{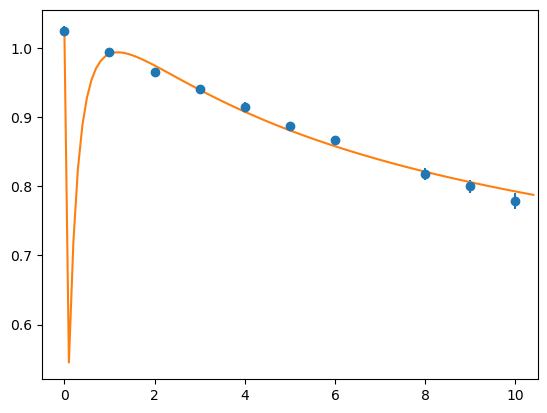}
\includegraphics[width=0.45\linewidth,clip]{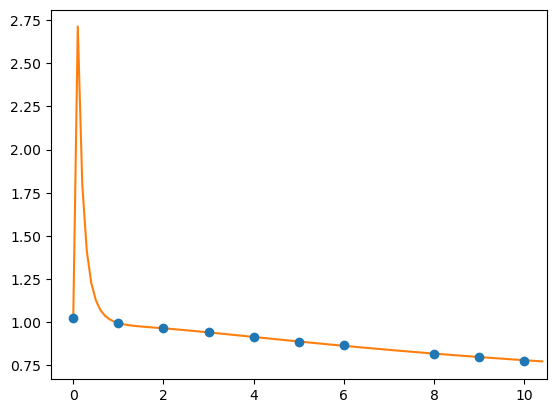}
\end{center}
\caption{
\label{fig:naiveF1}
Fits with \(k_{\rm max} = 3\) (left pane) and 4 (right pane) using a naive \(Q^2 \mapsto z\) mapping with \(t_0 = 0\) to the isovector vector form factor, \(F_1\), calculated on the RBC+UKQCD ''48I'' ensemble.
The large gap in \(z\) between \(Q^2=0\) and one lattice unit allows unphysical kinks in the fits.
}
\end{figure}

Naive fits to the isovector vector form factor, \(F_1\), calculated on the ``48I'' ensemble \cite{Ohta:2022csu,RBC:2014ntl} with a source-sink separation of 8 lattice units, for \(t_0 = 0\) and \(k_{\rm max} = 3\) and  4 are presented in Fig.\ \ref{fig:naiveF1}.
The large gap in \(z\) between \(Q^2=0\) and one lattice unit allows unphysical kinks in the fits.

\begin{figure}[b]
\begin{center}
\includegraphics[width=0.45\linewidth]{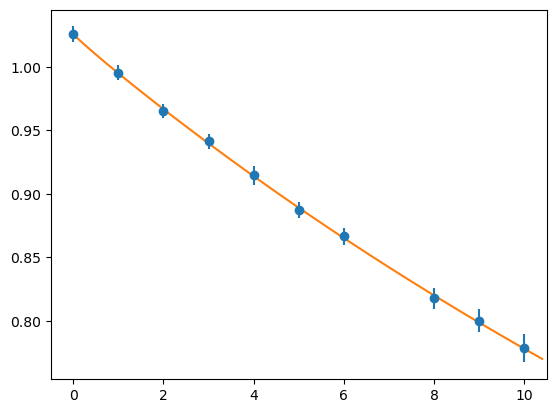}
\includegraphics[width=0.45\linewidth]{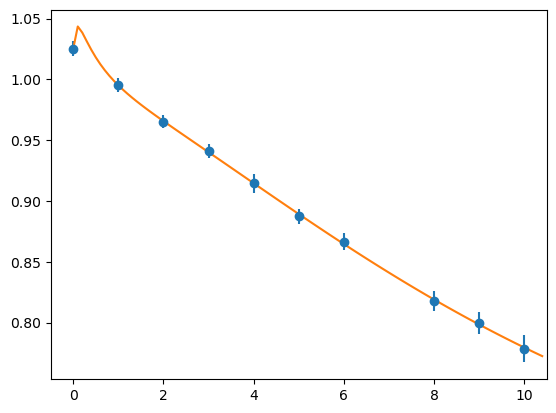}
\end{center}
\caption{
\label{fig:Q2=10F1}
A better fits with \(k_{\rm max} = 4\) (center pane) and 5 (right pane) using  \(Q^2 \mapsto z\) mapping with \(t_0 = -Q^2_{\rm max}\) to the isovector vector form factor, \(F_1\), calculated on the RBC+UKQCD ''48I'' ensemble.
The smaller gap in \(z\) between \(Q^2=0\) and 1 lattice units now allows a tamed fit with \(k_{\rm max} = 4\) to give a ``much improved'' estimate of  \(\langle r_1^2 \rangle \sim ({\rm 0.45 fm})^2\).
However, the estimate changes the sign when \(k_{\rm max} \) is increased to 5.
}
\end{figure}

This can be improved by using \(t_0=-Q^2_{\rm max}\), as presented in Fig.\ \ref{fig:Q2=10F1}.
The smaller gap in \(z\) between \(Q^2=0\) and 1 lattice units now allows a tamed fit with \(k_{\rm max} = 4\) to give a ``much improved'' estimate of  \(\langle r_1^2 \rangle \sim ({\rm 0.45 fm})^2\).
However, with \(k_{\rm max} = 5\) (right) the kink from the gap in \(z\) returns and changes the sign of \(\langle r_1^2 \rangle\).

These results point to a need for better constraining the polynomial coefficients.

\section{Constrained bounded-\(z\) polynomial expansion}

Indeed  a useful constraint arises from QCD: at large \(Q^2\) the form factors should fall at least as fast as \(1/Q^4\) \cite{Lepage:1980fj}: 
\begin{equation}
Q^nF(Q^2)\rightarrow 0,
\end{equation}
for \(n=0\), 1, 2, and 3.
Since \(\lim_{Q^2 \rightarrow \infty} z = 1\), these are equivalent with 
\begin{equation}
\frac{d^nF(z)}{dz^n}\, \biggl\rvert_{z=1} = 0, 
\end{equation}
for \(n = 0\), 1, 2, and 3.
These constrain the polynomial form to
\begin{equation}
F(z) = (1-z)^4 B(z)
\end{equation}
with arbitrary polynomial \(B(z)\), because
\begin{enumerate}
\item \(n=0\) leads to \(F(z) = (1-z)E(z)\),
\item \(n=1\) leads to \(E(z) = (1-z)D(z)\),
\item \(n=2\) leads to \(D(z) = (1-z)C(z)\),
\item \(n=3\) leads to \(C(z) = (1-z)B(z)\).
\end{enumerate}
These are, of course, equivalent to the more conventional `sum rules' \cite{Lee:2015jqa} in the literature.

\begin{figure}[t]
\begin{center}
\includegraphics[width=0.45\linewidth]{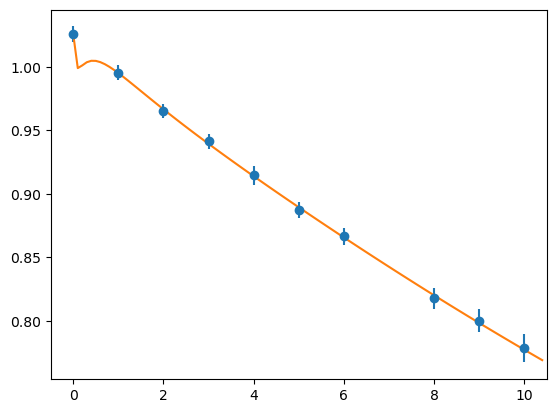}
\includegraphics[width=0.45\linewidth]{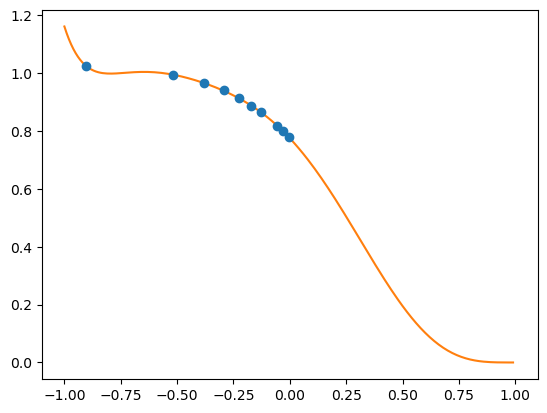}
\caption{
\label{fig:constrained8}
Constrained fit with \(k_{\rm max} =8\) to the isovector vector form factor, \(F_1\), plotted against \(Q^2\) in lattice units (left pane) and against bounded-\(z\) (right pane).
Good \(\chi^2\) per degree of freedom \(< 1\) requires this order, \(k_{\rm max} =8\), but also brings the kink from the gap in \(z\) between \(Q^2=0\) and 1.
This results in a rather unphysical, yet positive, estimate for \(\langle r_1^2 \rangle\).
}
\end{center}
\end{figure}

\begin{figure}[b]
\begin{center}
\includegraphics[width=0.45\linewidth]{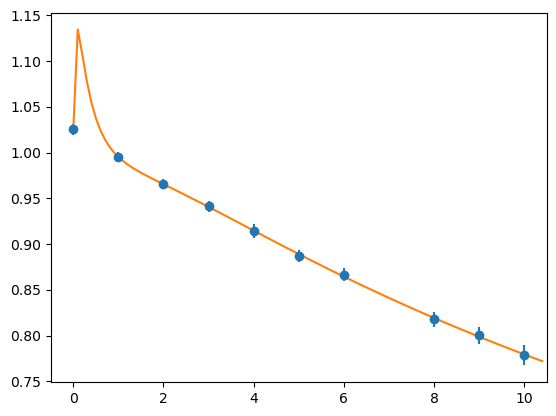}
\includegraphics[width=0.45\linewidth]{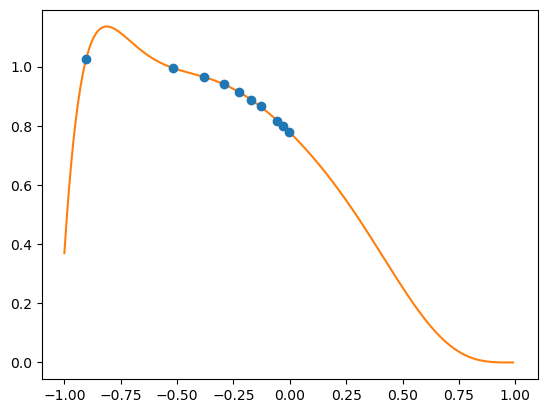}
\caption{
\label{fig:constrained9}
Constrained fit with \(k_{\rm max} =9\) to the isovector vector form factor, \(F_1\), plotted against \(Q^2\) in lattice units (left pane) and against bounded-\(z\) (right pane).
This does not improve on \(k_{\rm max} =8\), but merely changes the sign of the estimated \(\langle r_1^2 \rangle\), again because of the kink from the gap in \(z\) between \(Q^2=0\) and 1.
}
\end{center}
\end{figure}

We found that good fits with \(\chi^2\) per degree of freedom \(< 1\) for the present nucleon isovector form factors usually require \((1-z)^4 \times\) fourth-order or higher polynomials (see Figs.\ \ref{fig:constrained8} and \ref{fig:constrained9}).
However those with \(k_{\rm max} = 8\) (left) and 9 (right) differ in \(\langle r_1^2\rangle\) signs.

These behaviors do not change as we vary the fitting range from all the ten \(Q^2\) points between 0 and 10 lattice units to a) \(0\le Q^2 \le 6\), or b) every other points \(Q^2 = 0, 2, 4, 6, 8,\) and 10.
However, removing the \(Q^2 = 0\) point makes the fits much less controlled toward \(Q^2 \rightarrow 0\).
Also, these behaviors do not change as we look at the other form factors, \(F_2\), \(F_A\), and \(F_P\), with all the source-sink separations of 8, 9, 10, 11, and 12 lattice units.
Nor do the behaviors change as we move \(t_0\) farther away, to two orders of magnitude larger.

\section{Conclusion}

We found the QCD-constrained bounded-\(z\) expansions for the present nucleon isovector form factors do not provide their shape parameters in agreement with those extracted by linear or dipole fits.
In particular, we observe the following:
\begin{enumerate}
\item the linear and dipole extractions as summarized in Table \ref{tab:summary}, which are driven by the calculated form factor values at \(Q^2=0\) and 1, and 1 and 2 in lattice units, do not seem to agree well with the respective experiments, and
\item the bounded-\(z\) polynomial fits we so far tried are not stable because of the large gap in \(z\) between  \(Q^2=0\) and 1.
\end{enumerate}
We note Bayesian priors \cite{Alexandrou:2018zdf,Flynn:2023qmi} may help to further constrain the bounded-\(z\) polynomial fits.
However, if such a prior works merely to thin the influence from large \(Q^2\), larger than 2 in lattice units,
using the smallest two \(Q^2\) points for linear extrapolations likely works better.

We need smaller momentum transfer units before the calculations can be compared with the experiment.
Lattice-QCD calculations with smaller momentum transfer units can be achieved by either larger volumes or different boundary conditions.

\section*{Acknowlegment}

The author thanks the LHP, RBC, and UKQCD collaboration members, particularly Sergey Syritsyn, for leading the 48I nucleon calculations and Andreas Juettner for enlightening discussions about bounded-\(z\) expansion.
The 48I ensemble was generated using the IBM Blue Gene/Q (BG/Q) ``Mira'' machines at the Argonne Leadership Class Facility (ALCF) provided under the Incite Program of the US DOE, on the ``DiRAC'' BG/Q system funded by the UK STFC in the Advanced Computing Facility at the University of Edinburgh, and on the BG/Q  machines at the Brookhaven National Laboratory.
The nucleon calculations were done using ALCF Mira.
The Japan Society partially supported the author for the Promotion of Sciences, Kakenhi grant 15K05064.
Part of the work was conducted while the author was affiliated with the RIKEN-BNL Research Center through March 31, 2021.

\bibliographystyle{apsrev4-2}
\bibliography{nucleon}

\end{document}